# Living Images: A Recursive Approach to Computing the Structural Beauty of Images or the Livingness of Space

Bin Jiang and Chris de Rijke

Faculty of Engineering and Sustainable Development, Division of GIScience
University of Gävle, SE-801 76 Gävle, Sweden
Email: bin.jiang|chris.de.rijke@hig.se

*In an organic environment, every place is unique, and the different places also cooperate, with no parts left over, to create a global whole - a whole which can be identified by everyone who is part of it.*

Christopher Alexander et al. (1975)

**Abstract**
According to Gestalt theory, any image is perceived subconsciously as a coherent structure (or whole) with two contrast substructures: figure and ground. The figure consists of numerous auto-generated substructures with an inherent hierarchy of far more smalls than larges. Through these substructures, the structural beauty of an image (L), or equivalently the livingness of space, can be computed by the multiplication of the number of substructures (S) and their inherent hierarchy (H). This definition implies that the more substructures something has, the more living or more structurally beautiful it is, and the higher hierarchy of the substructures, the more living or more structurally beautiful. This is the non-recursive approach to the structural beauty of images or the livingness of space. In this paper we develop a recursive approach, which derives all substructures of an image (instead of its figure) and continues the deriving process for those decomposable substructures until none of them are decomposable. All of the substructures derived at different iterations (or recursive levels) together constitute a living structure; hence the notion of living images. We have applied the recursive approach to a set of images that have been previously studied in the literature and found that (1) the number of substructures of an image is far lower (3 percent on average) than the number of pixels and the centroids of the substructures can effectively capture the skeleton or saliency of the image; (2) all the images have the recursive levels more than three, indicating that they are indeed living images; (3) no more than 2 percent of the substructures are decomposable, implying that a vast amount of the substructures are not decomposable; (4) structural beauty can be well measured by the recursively defined substructures, as well as their decomposable subsets. Despite a slightly higher computational cost, the recursive approach is proved to be more robust than the non-recursive approach. The recursive approach and the non-recursive approach both provide a powerful means to study the livingness or vitality of space in cities and communities.

## 1. Introduction
All space has some degree of livingness in it, according to its structure and arrangement (Alexander 2002–2005), so the livingness is commonly sensed in our surrounding such as rooms, gardens, buildings, streets, and cities, as well as in tiny ornaments. The livingness sounds like a kind of human experience of space or a sense of place attachment (Tuan 1977, Goodchild and Li 2012), synonymous with the vitality or organized complexity (Jacobs 1961), and the imageability or legibility (Lynch 1960). Unlike these concepts, however, the livingness is defined mathematically through the underlying living structure. The living structure is a mathematical structure with an inherent hierarchy (see Section 2 for an introduction), which can trigger the feeling of livingness in the human mind and heart. The inherent hierarchy of living structure is commonly recognized in a series of urban and geographic theories such as the central place theory (Christaller 1933, 1966), space syntax (Hillier and Hanson 1984, Hillier 1996), and the fractal cities (Batty and Longley 1994). The hierarchy may reflect spatial heterogeneity, one of the two spatial properties, the other being spatial dependence (Goodchild 2004, Anselin 1989, Tobler 1970). As shown later in Section 2, the hierarchy or spatial heterogeneity is better characterized by the recurring notion of far more smalls than larges across different levels of scale.

Any image is perceived by human eyes subconsciously as a coherent structure (or whole) with two large, contrasting substructures: figure and ground (Koffka 1936, Rubin 1921). The figure, which is also called the foreground, constitutes the focus of the visual field, while the ground is the background. The figure is a living structure, for it can be decomposed into many substructures with an inherent hierarchy of far more smalls than larges. The substructures are auto-generated segments (or sets of pixels) out of a gray-scale image by vectorizing the individual sets of pixels that are darker (or lighter) than the average pixel. There are far more small substructures than large ones across the hierarchy or the different levels of scale, yet the substructures on each of the hierarchy are more or less similar in size. It is essentially the recurring notion of far more small substructures than large ones that triggers a sense of livingness in the human mind and heart (Jiang 2019). This sense of livingness is called structural beauty (Jiang and De Rijke 2021) and it is shared among people, and even different peoples, regardless of their cultures, gender, and races. Thus, the livingness (L) or structural beauty is defined by the derived substructures, or more specifically the multiplication of their number (S) and their inherent hierarchy (H); that is, $L = S \times H$. Instead of working with the figure, we, in the present paper, work directly with the image itself and derive its substructures, and the substructures of the decomposable substructures recursively until all substructures are no longer decomposable. All the substructures at different iterations (or recursive levels) together constitute a coherent whole or a living structure: hence the notion of living images, the central theme of this paper.

In this paper we consider an image – or space in general – to be a living structure that is composed of recursively defined substructures. This is a holistic view of perceiving an image or space as a coherent whole, so it differs fundamentally from conventional thinking (e.g., Umbaugh 2017, Davies 2017). Conventional image understanding tends to identify a few features or objects that can be named by words or recognizable by human eyes – so-called computer vision. For example, a human face image consists of numerous substructures with far more smalls than large ones, but our natural language can only name certain features, such as the eyes, the nose, the mouth, the ears, and the hair. In other words, a vast majority of substructures cannot be named by words. In general terms, a gray-scale image can be decomposed, around the average pixel value m1, into dark pixels (darker than m1) which may be called the figure (say, 52 percent) and light pixels (lighter than m1), which may be called the ground (for example, 48 percent). Although the figure (or the dark pixels) is perceived as a whole, it consists of numerous substructures with an inherent *hierarchy* of far more smalls than larges. Interestingly, the figure can be further decomposed in a recursive manner, around the average pixel value (mi) of the figure, into dark and light pixels, leading to numerous substructures with far more smalls than larges. This decomposition process is referred to here as *recursion*.

We have mentioned two concepts so far: hierarchy and recursion. The hierarchy refers to the recurring notion of far more small substructures than large ones on each of the recursion, whereas the recursion refers to the decomposition process for those substructures that are decomposable at different levels of the recursion. Here we use a parable to further clarify these two concepts. Imagine a tree with five hierarchical levels: one trunk, two limbs, eight branches, 24 twigs, and 96 leaves. According to the above mentioned formula, $L = S \times H$, the degree of structural beauty is calculated through the multiplication of substructures (1 + 2 + 8 + 24 + 96 = 134) and their hierarchy (5); that is, $134 \times 5 = 670$. We know that the leaves have their own textures that have a similar hierarchy to the tree itself (for example, five levels). This means that each of the leaves can be decomposed into five hierarchical levels (Jiang and Huang 2021). Thus, there are two levels of recursion, and 96 decomposable substructures (or leaves), which means there is an alternative way of calculating the degree of structural beauty is $96 \times 2 = 192$ (to be introduced as a new formula in Section 3.2). It is this new insight about the recursive nature of substructures that motivated us to develop this paper.

The contribution of this paper lies in the living structure perspective, a holistic and comprehensive approach to measuring the structural beauty of images or the livingness of space in a recursive manner. More specifically, there are four major findings from this study. First, all images are living images that have recursive levels more than four. Second, the centroids of the recursively defined substructures effectively capture the skeleton or saliency of the images, and the number of centroids (or substructures) is far fewer than the number of pixels. Third, among the derived substructures, no more than 2 percent are decomposable. Fourth, not only substructures but also their decomposable subsets can be used to measure the structural beauty of images or the livingness of space.

The remainder of this paper is structured as follows. Section 2 introduces the concept of living structure and explains why one structure is more living or structurally beautiful than another, not as an idiosyncratic opinion, but as a matter of measurable fact. Section 3 uses a working example to illustrate a recursive approach to computing the livingness of space or structural beauty of images. Section 4 reports our case studies for verification of the recursive approach and major findings out of the case studies. In Section 5 we further discuss on the livingness of space in terms of related works, its application, and implications on geography and even beyond. Finally, in Section 6 we draw a conclusion and point to future work.

## 2. Living structure, the degree of structural beauty or livingness, and two laws

Living structure is what underlies the notion of structural beauty or the livingness of space. As mentioned above, the degree of structural beauty of an image (L) is determined by its substructures, or their number (S) and their hierarchy (H) to be more precise. In other words, the more substructures an image has, the more living or more structurally beautiful it is, and the higher hierarchy of the substructures, the more living or more structurally beautiful the image is. This rule about livingness or structural beauty was distilled from the 15 properties (cf. Figure 1), originated from the 253 patterns (Alexander et al. 1977). These properties recur in natural things – naturally occurring things – as well as in human-made things. Those natural and human-made things are genuinely beautiful and can trigger a sense of livingness in the human mind and heart (Alexander and Carey 1968, Alexander 1993, 2002–2005). In this section, we first compare two drawings – as two living structures – in terms of their livingness through calculating the L score. Eventually, the one with the higher L score is more living or more structurally beautiful. These two working examples help us to understand why one structure is more living than another, and importantly two fundamental laws of living structure.

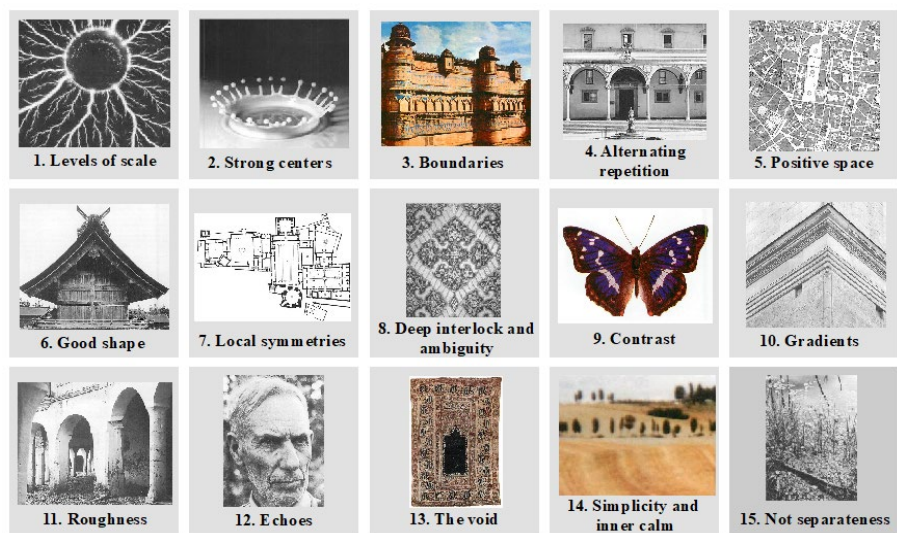

Figure 1: (Color online) Alexander's fifteen properties that exist in natural and human-made things (Note: These 15 properties (Alexander 2002–2005) were distilled from the 253 patterns (Alexander et al. 1977).)

Two drawings are conventionally seen as two sets of four squares and four octagons, respectively (Panels a and b of Figure 2). Instead of this static or LEGO-like assembly view, we consider these two drawings to be evolved from an empty square of size one through the three iterations (Panel d). The two drawings have a clear figure distinguished from the white background: the four squares and the four octagons, respectively. The four squares (or four substructures in general, S = 4) have two hierarchical levels (H = 2), so the four squares as a whole have structural beauty of $4 \times 2 = 8$ (Panel c). Alternatively seen from the perspective of 16 sides, the structural beauty is $16 \times 2 = 32$. For the right drawing (Panel b), there are four octagons (S = 4) that are defined at three hierarchical levels: 1, 1/3, and 1/9 (H = 3), which means that structural beauty is $4 \times 3 = 12$. Alternatively, there are 32 edges of the four octagons defined at the three hierarchical levels, so structural beauty is $32 \times 3 = 96$. From the calculation, there is little doubt that drawing (b) is more living or more structurally beautiful than drawing (a). The above computations are shown in Panel c.

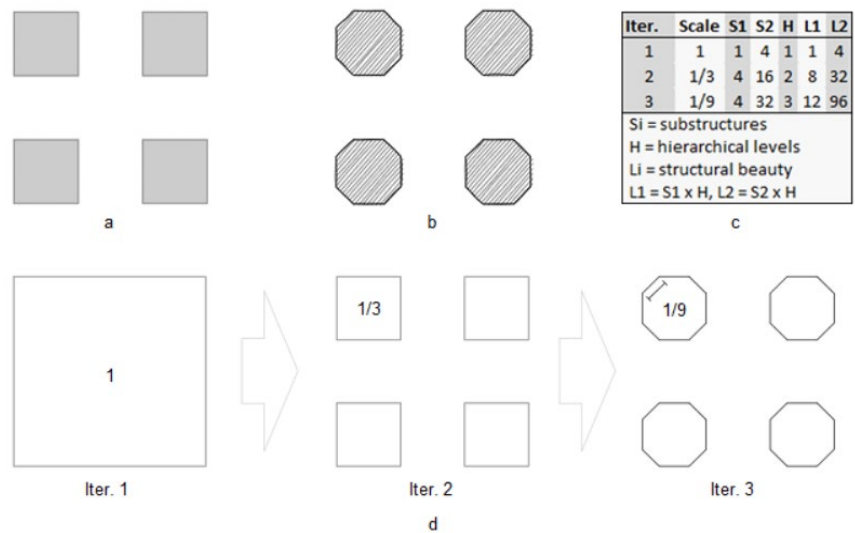

Figure 2: Illustration on why one structure is less living than another
(Note: The drawing (Panel a) is less living or less beautiful structurally than the other drawing (Panel b). Instead of seeing each of the structures (Panels a and b) as a collection of the substructures (squares or octagons), we see them to be transformed from the empty square in a step-by-step fashion (Panel d). In Panel d, the empty square to the left has scale 1 and is decomposed into the four squares of scale 1/3, and further to the four octagons with a scale of about 1/9. During the transformation, there is twice a recurring notion of far more newborns (or newly generated substructures) than old ones: the first time for the structure of the left drawing (Panel a) and the second time for that of the right drawing (Panel b). Seen from the point of view of transformation, the left drawing (Panel a) is less living than the right drawing (Panel b) because the left possesses few substructures with a lower hierarchy, as shown in Panel c. Alternatively, the right drawing (Panel b) can be seen to be transformed directly from the left drawing (Panel a) by cutting out the 16 corners, so it becomes more differentiated. Through the calculation, there is little doubt that the left drawing is less living than the right one.)

Table 1: Two complementary laws of living structure

| Scaling law | Tobler's law |
|---|---|
| Far more small substructures than large ones across all scales | More or less similar substructures on each of scales |
| Disproportional ratio of smalls to larges (80/20) | Proportional ratio of smalls to larges (50/50) |
| Scale-free or scaling globally | With a characteristic scale locally |
| Pareto distribution or a heavy-tailed distribution | Gauss-like distribution |
| Spatial heterogeneity or interdependence | Spatial homogeneity or dependence |
| Complex and non-equilibrium phenomena | Simple and equilibrium phenomena |

Having shown the two examples of living structure, let's see how it can be defined formally. Living structure is a physical phenomenon pervasively existing in our surroundings and is a mathematical structure that consists of or is transformed in a step-by-step fashion into numerous substructures with an inherent hierarchy. Across different levels of the hierarchy there are far more small substructures than large ones, yet on each level of the hierarchy substructures are more or less similar. These two notions – far more smalls than larges, and more or less similar – underlie the scaling law (Jiang 2015) and Tobler's law (1970), respectively. These two laws are complementary rather than contradictory to each other (Table 1) to characterize the various kinds of living structure in natural and man-made things. These two laws will be further illustrated in the next section.

Strictly speaking, neither of the two drawings is a living structure because they both violate Tobler's law. The four squares or four octagons are precisely the same rather than more or less similar. In addition, the left drawing violates the scaling law, which states that the notion of far more smalls than larges recurs at least twice or with the hierarchy being three. Nevertheless, the two drawings are simple enough to illustrate various concepts such as living structure, substructures, hierarchy, two fundamental laws, and the livingness of space.

### 3. A recursive approach to computing the structural beauty of images
In this section, we introduce a recursive approach to the structural beauty of images by extending the substructures to recursive ones. Before that, we will also illustrate the head/tail breaks (Jiang 2013) as a recursive function for deriving the underlying living structure or the inherent hierarchy of a dataset. The dataset contains 100 numbers as a working example to mimic the rank–size distribution: the largest size is about twice as large as the second largest, approximately three times as large as the third largest, and so on (Zipf 1949). This working example also helps to illustrate the two laws of living structure.

### 3.1 Head/tail breaks and two laws of living structure
Head/tail breaks was initially developed as a classification scheme for a dataset with a heavy-tailed distribution. It is a de facto recursive function for deriving the inherent hierarchy of a dataset from the bottom up. That is, the dataset itself determines the inherent hierarchy of classes without any imposed criteria. Let us use the dataset containing 100 numbers [1, 1/2, 1/3, …, 1/100] (Figure 3) to illustrate how the dataset can be decomposed into the head for those greater than the average and the tail for those less than the average, and then proceed recursively or iteratively for the head, the head of the head, and so on. All tails and the last head constitute individual classes or hierarchical levels.

To be specific, the average of the 100 numbers is approximately 0.05, and it divides the dataset into two subsets: the head for those greater than the average [1, 1/2, 1/3,…, 1/19] and the tail for those less than the average [1/20, 1/5, …, 1/100]. For the 19 numbers in the head subset, the average is about 0.19, and it divides the head subset into the head [1, 1/2, 1/3,1/4, 1/5] and the tail [1/6, 1/7,…,1/19]. For the five numbers in the latest head subset, the average is about 0.46, and it divides the latest head subset into the head [1, 1/2] and the tail [1/3,1/4, 1/5]. Thus, there are four classes or hierarchical levels for the data: [1, 1/2], [1/3, 1/4, 1/5], [1/6, 1/7,…, 1/19], [1/20, 1/21, …, 1/100]. The data can be conceived to be composed of the head of the head of the head of an iterative system: [1, 1/2], [1, 1/2, 1/3, 1/4, 1/5], [1, 1/2, 1/3, …, 1/19], [1, 1/2, 1/3, …, 1/100].

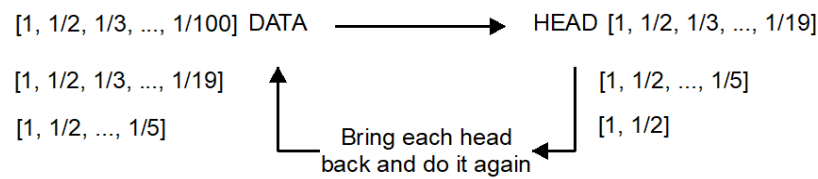

Figure 3: Illustration of head/tail breaks as a recursive function
(Note: The dataset is divided around the average into the head for those larger than the average and the tail for those smaller than the average. The head is brought back to do the same division again and again recursively until a certain threshold is met. The tails and the last head constitute individual classes in the iterative order: [1/20, 1/21, …, 1/100], [1/6, 1/7, …, 1/19], [1/3, 1/4, 1/5], and [1, 1/2]. The notion of far more small numbers than large ones for the dataset recurs three times across the hierarchy, meeting the scaling law. On the other hand, there are more or less similar numbers on each of the hierarchy, meeting Tobler's law.)

The working example can further illustrate the two laws of living structure (Table 1). The dataset clearly has the recurring notion of far more smalls than larges, so it meets the scaling law. More specifically, the notion of far more smalls than larges recur three times, so with four different levels of scale (or hierarchy). On the other hand, the numbers on each level of the hierarchy (or scale) are more or less similar in size, so the dataset meets Tobler's law. Among the two laws, the scaling law is available across the hierarchy while Tobler's law on each level of the hierarchy. More importantly, the scaling law and Tobler's law imply complex and non-equilibrium character globally, and simple and equilibrium character locally, respectively.

## 3.2 The recursive approach

Applying the head/tail breaks to a gray-scale image will help get the substructures of the image. More specifically, we first determine the figure or the ground of the image depending on the percentage of the dark or light pixels. The high percentage is usually referred to as the ground, while the low percentage is referred to as the figure (Robin 1921). The first dichotomy process helps differentiate the figure from the image and helps derive the first round of substructures. To illustrate, let us take the weather-beaten face image as an example (Figure 4). It consists of 390 by 672 (262,080) pixels, each of which has a gray scale between 26 and 254. The average pixel value of those 262K pixels is 163, which partitions all the pixels into two groups: 120,036 dark pixels (46 percent of the total) as the figure, and 142,044 light pixels (54 percent) – the so-called dichotomy. The dark pixels are set to black, while the light pixels are set to white to create a binary image (binarization), which is further vectorized into a living structure consisting of numerous substructures. To this point, the illustration fulfills the first iteration, as shown in Panel a of Figure 4. As a reminder, the first iteration or the first level of recursion that starts from the image is NOT identical to the no-recursive approach (Jiang and De Rijke 2021), which is based on the figure of the image for deriving all non-recursively defined substructures.

Table 2: Statistics of recursively defined substructures and their degrees of life
(Note: This table supplements Figure 4 to show the underlying statistics of substructures, including decomposable and undecomposable, and the degree of structural beauty (LR). I = Iteration, D = Decomposable, S = Substructures, H = Hierarchy of the substructures, U = Undecomposable, and % = percentage of D to S. The underlined numbers of substructures are derived from multiple decomposable substructures, so the corresponding H are calculated from LR/S.)

| I | D | S | H | U | % | LR |
|---|---|---|---|---|---|---|
| 1 | 1 | 768 | 4.0 | 757 | 1.4% | 3,072 |
| 2 | 11 | 1,856 | 3.9 | 1,836 | 3.0% | 7,264 |
| 3 | 20 | 1,206 | 4.3 | 1,184 | 2.3% | 5,161 |
| 4 | 22 | 598 | 3.6 | 592 | 1.8% | 2,141 |
| 5 | 6 | 112 | 3.4 | 109 | 1.0% | 380 |
| 6 | 3 | 26 | 3.0 | 25 | 0.0% | 78 |
| Σ | 63 | 4,566 | | 4,503 | | 18,096 |

The recursive approach starts with the first iteration or the first level of recursion (as mentioned above) and use the first iteration substructures to clip the original image, and then continue what was done in the iteration again and again, until all decomposable substructures are decomposed. As shown in Figure 4, the first iteration leads to 768 substructures (Panel b) and their centroids (Panel c1), 11 of which are decomposable, leading to 1,856 substructures (whose centroids in Panel c2), 20 of which are decomposable, leading to 1,206 substructures (whose centroids in Panel c3), 22 of which are decomposable, leading to 598 substructures (whose centroids in Panel c4), six of which are decomposable, leading to 112 substructures (whose centroids in Panel c5), three of which are decomposable, leading to 26 substructures (whose centroids in Panel c6). Table 2 provides detailed statistics for the recursive process. All the centroids of the recursively derived substructures effectively capture the skeleton or saliency of the image (Panel d). The centroids of the 63 = 1 + 11 + 20 + 22 + 6 + 3 decomposable substructures are shown in Panel e.

According to the previous work by Jiang and De Rijke (2021), the livingness (L) of space or structural beauty of images for the non-recursive approach is formally defined by

$$L = S \times H \quad [1]$$

where S and H denote the number and the hierarchy of substructures, respectively. This definition implies that the more substructures an image has, the more beautiful it is, and the higher the hierarchy of the substructures of the image, the more beautiful it is.

This formula is transformed into the following format for multiple levels of recursion,

$$LR = \sum_{i=1}^{D} S_i \times H_i \quad [2]$$

where D indicates the total number of decomposable substructures (including the image itself), and i the individual decomposable substructures.

For the weather-beaten face image, the first decomposable substructure (at the first iteration) is the image itself and it has 768 substructures defined at four hierarchical levels, so $LR_1 = 768 \times 4 = 3,072$; for the last three decomposable substructures (at the sixth iteration), so $LR_{61-63} = 7 \times 3 + 10 \times 3 + 9 \times 3 = 78$. In Table 2, we divided the LR score according to individual iterations, so the hierarchy (H) is derived from H = LR/S, which is why H is not integral, except for the first and the last iterations.

As an alternative measure, we define structural beauty (V) based on the decomposable substructures (D) and their iterations or the levels of recursion (I):

$$V = D \times I \quad [3]$$

As an example, the weather-beaten face image has 63 decomposable substructures derived at the six levels of recursion (Table 2), so $V = 63 \times 6 = 378$.

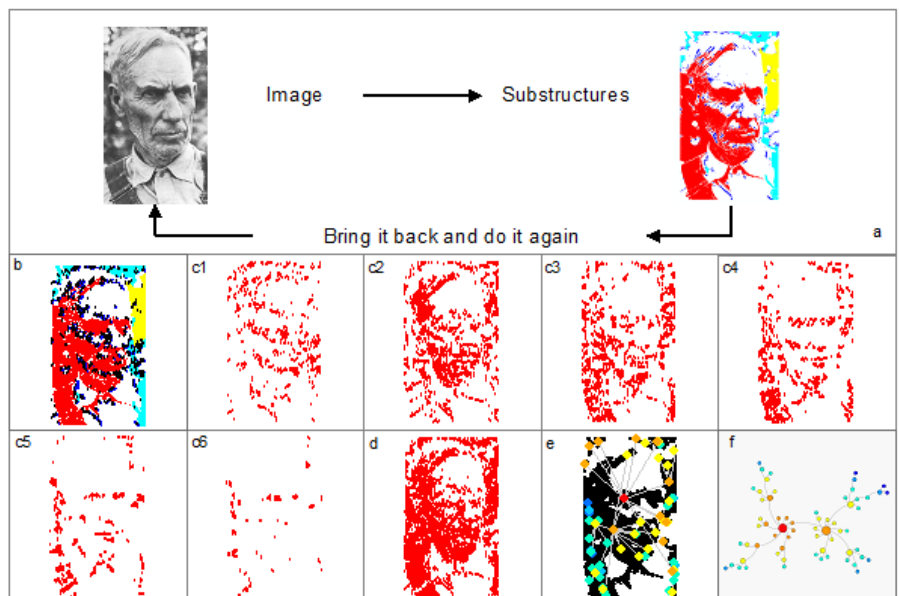

Figure 4: (Color online) Illustration of the recursive approach to the structural beauty of images (Note: The weather-beaten face image is decomposed into numerous substructures – recursively – with an inherent hierarchy of far more smalls than larges (a) where the four colors indicate the four levels of the hierarchy. More specifically, the image – as a whole – is binarized around the average pixel and further vectorized into 768 substructures, which are converted into 768 centroids represented by the black dots (b). This process continues recursively for the decomposable

substructures, resulting in 768, 1,856, 1,206, 598, 112, and 26 substructures, respectively, at different levels of the recursion, and their centroids are shown as red dots (c1–c6). Overlapping all the six levels of the 4,566 (= 768 + 1,856 + 1,206 + 598 + 112 + 26) centroids together represents the skeleton of the image (d). Among the 4,566 substructures, only 62 (= 11 + 20 + 22 + 6 + 3) are decomposable substructures and their centroids (e), and the 63 centroids are visualized according to their degree of connectivity (both in- and out-links) by the dot sizes, with colors indicating their levels of recursion (f).)

A short note on the concept of substructures follows. Conventional image understanding and computer vision tends to identify individual objects that human beings can recognize in a human face image, such as the eyes, nose, and mouth, but hundreds and thousands of objects we cannot find proper words to name. Instead, the concept of substructures is defined naturally or organically by all pixels. In other words, all the pixels collectively determine an average pixel value that is used to delineate individual substructures, a kind of wisdom of crowds thinking (Surowiecki 2004). More importantly, unlike objects that are fragmented pieces, the substructures constitute a coherent whole. It is the concept of whole or its substructures that make the approach unique and different from the conventional image understanding.

In the following case studies, we will examine, among other things, (1) whether the structural beauty calculated from the recursively defined substructures can help differentiate two images using Formula [2], and (2) whether the decomposable substructures are able to differentiate two images based on Formula [3].

## 4. Case studies

We applied the recursive approach to the eight pairs of images to verify whether the recursive approach is better or more robust than the non-recursive approach. The same set of images had been used to verify the non-recursive approach previously (Jiang and de Rijke 2021), so they were convenient for verification of the recursive approach and for comparing the non-recursive and recursive approaches. The first four pairs had previously been studied by Alexander (2002–2005), who used the 15 properties to examine their livingness, so they are with ground truth on their livingness or structural beauty. In addition to the verification and comparison, we found that (1) the number of the recursively defined substructures of an image is far smaller (3 percent on average) than the number of pixels, and the centroids of these substructures can well capture the skeleton or saliency of the image; (2) all the images have the recursive levels more than four, indicating that they are indeed living images; and (3) no more than 2 percent of the substructures are decomposable, but they can well characterize the structural beauty.

### 4.1 Verification of the recursive approach

The eight pairs are put into two groups: the building group (P1–P4) and the mixed group (P5–P8) (Figure 5). For every pair of images, the left is supposed to be more living or more structurally beautiful than the right, according to the non-recursive approach (Jiang and de Rijke 2021). The new approach can duplicate the same result (Table 3), where both columns (L and V) indicate that the left-hand image has a higher score than the right-hand image. Our primary goal was to differentiate between the two images to see whether the left image's score is higher than that of the right image. This is indeed true; see Columns L (calculated from all substructures) and V (calculated from only decomposable substructures). All the images to the left are more living or more structurally beautiful than those to the right, and all the images are living images, indicated by at least four iterations or four levels of recursion.

### 4.2 Centroids of the substructures capture the skeleton of the images

The centroids of the substructures capture very well the skeleton or saliency of the images. This is shown in Figure 5 as living structures in red, whose numbers are shown in column "substructures" in Table 3. They account for an average of 3 percent of the pixels of an image. We conjecture that these centroids are what is captured by a painter while he/she is drawing a sketch of the image. To demonstrate how the centroids capture the skeleton or saliency, Figure 6 shows three enlarged pairs from Figure 5. The centroids of the substructures are representative of the corresponding images. This is probably the reason that structural beauty calculated from these substructures (or their centroids) can effectively differentiate the two images.

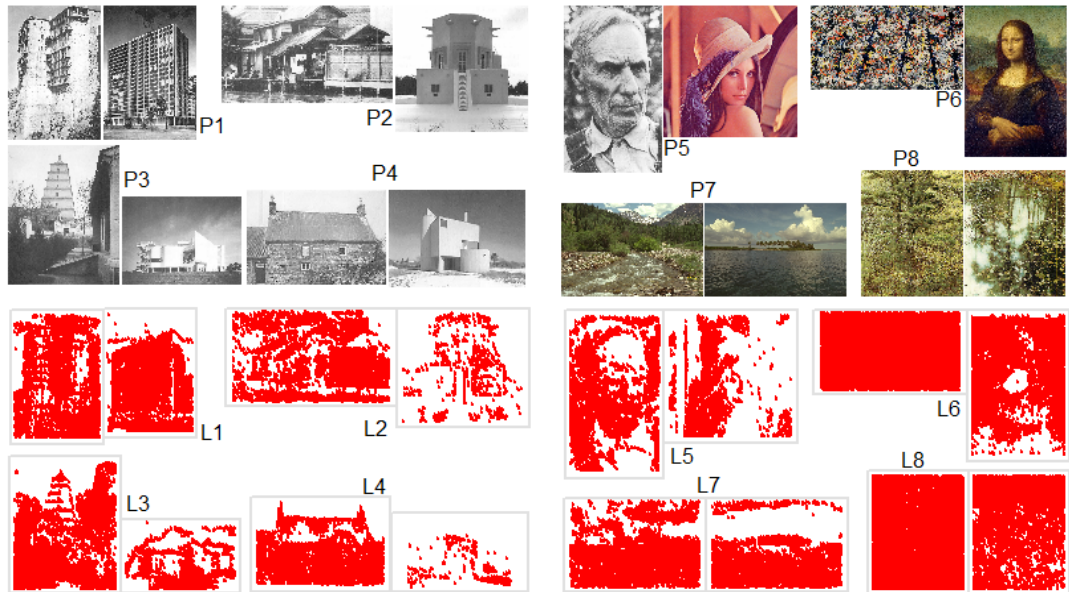

Figure 5: (Color online) The eight pairs of images and centroids of their substructures
(Note: The first four groups of images are with ground truth about which one is more living or beautiful than the other; that is, traditional buildings are more living than the modernist counterparts (Alexander 2002–2005). In the second four groups, according to the previous study (Jiang and De Rijke 2021), the left is more beautiful than the right for every pair. Image source: P1–P4, P5(left), P8 from Alexander (2002–2005), P5 (right), P7 from the image processing community, and P6 from Wikipedia.)

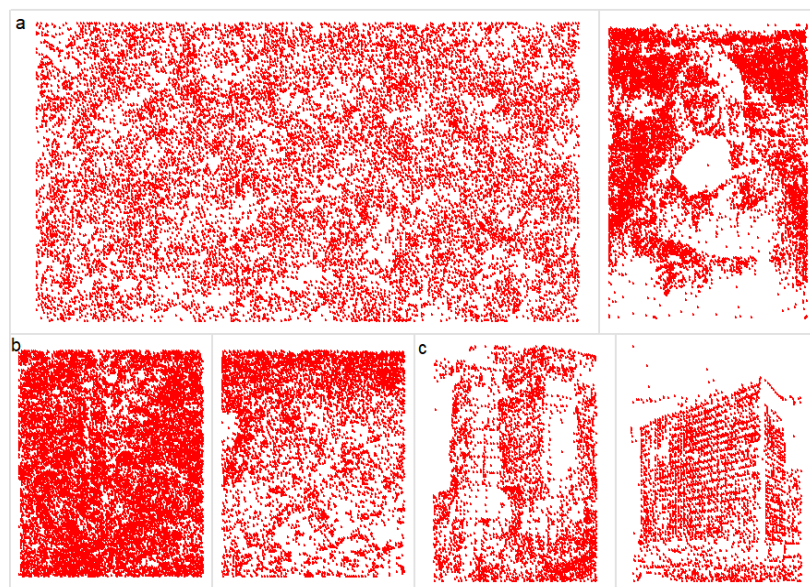

Figure 6: (Color online) Zoom-in view of the centroids of the substructures
(Note: Enlarged view of the three pairs of images as shown in Figure 5 for L6 (a), L8 (b) and L1 (c). For the images where figure and ground are very clearcut, like the two buildings and the Mona Lisa, the skeleton or saliency of the image is very striking, while for the images where figure and ground are fuzzy, the skeleton or saliency is not striking.)

Table 3: Structural beauty calculated from substructures and decomposable substructures
(Note: % (pixels) = Number/262K, % (centroids) = Number/all substructures, L = degree of structural beauty based on all substructures, Ranking(L) = Rank order according to L, V = another measure for the degree of structural beauty (vitality or vibrancy) based on the decomposable substructures, Ranking(V) = Rank order according to V.)

| Image pair | Image name | Substructures Number | % (pixels) | Decomposable Number | % (centroids) | Iteration | Degree of beauty LR | Ranking (LR) | V | Ranking(V) |
|---|---|---|---|---|---|---|---|---|---|---|
| P1 | Greek Monastery | 4,007 | 1.5% | 64 | 1.6% | 5 | 17,869 | 9 | 320 | 10 |
|    | Detroit Appartments | 3,324 | 1.3% | 24 | 0.7% | 4 | 16,652 | 10 | 96 | 12 |
| P2 | Slum | 4,492 | 1.7% | 86 | 1.9% | 5 | 18,264 | 7 | 430 | 6 |
|    | Postmodern Façade | 712 | 0.3% | 12 | 1.7% | 4 | 2,387 | 15 | 48 | 13 |
| P3 | The Tower of the Wild Goose | 5,998 | 2.3% | 46 | 0.8% | 7 | 23,174 | 6 | 322 | 9 |
|    | The X House | 2,304 | 0.9% | 16 | 0.7% | 4 | 7,144 | 14 | 64 | 14 |
| P4 | Traditional House | 14,592 | 5.6% | 30 | 0.2% | 5 | 14,592 | 11 | 150 | 11 |
|    | Postmodern House | 309 | 0.1% | 4 | 1.3% | 4 | 1,030 | 16 | 16 | 15 |
| P5 | Weatherbeaten Face | 4,566 | 1.7% | 63 | 1.4% | 6 | 18,096 | 8 | 378 | 7 |
|    | Lena Face | 2,931 | 1.1% | 56 | 1.9% | 6 | 11,463 | 13 | 336 | 8 |
| P6 | Blue Poles | 16,230 | 6.2% | 250 | 1.5% | 5 | 81,487 | 1 | 1,250 | 1 |
|    | Mona Lisa | 10,466 | 4.0% | 69 | 0.7% | 6 | 40,773 | 3 | 570 | 3 |
| P7 | Kodak Mountains | 6,005 | 2.3% | 74 | 1.2% | 6 | 29,297 | 4 | 444 | 5 |
|    | Kodak Island | 3,597 | 1.4% | 46 | 1.3% | 7 | 12,971 | 12 | 322 | 9 |
| P8 | Woods | 13,010 | 5.0% | 163 | 1.3% | 5 | 67,499 | 2 | 815 | 2 |
|    | Pond | 6,018 | 2.3% | 87 | 1.4% | 6 | 26,527 | 5 | 522 | 4 |

Table 4: Decomposed version of structural beauty (L) in comparison with the non-recursive approach
(Note: This table supplements Table 3 on column (L). The non-recursive approach is not correspondent to L0, as the former applies the figure of the image, while the latter applies to the image.)

| Image pair | Image name | Recursive approach LR0 | LR1 | LR2 | LR3 | LR4 | LR5 | LR6 | Total | Non recursive S | H | L |
|---|---|---|---|---|---|---|---|---|---|---|---|---|
| P1 | Greek Monastery | 7,585 | 3,872 | 4,899 | 1,441 | 72 | | | 17,869 | 6,248 | 6 | 37,488 |
|    | Detroit Appartments | 1,428 | 7,141 | 7,482 | 601 | | | | 16,652 | 7,241 | 4 | 28,964 |
| P2 | Slum | 6,150 | 5,532 | 4,086 | 2,024 | 472 | | | 18,264 | 3,799 | 6 | 22,794 |
|    | Postmodern Façade | 266 | 1,432 | 668 | 21 | | | | 2,387 | 362 | 4 | 1,448 |
| P3 | The Tower of the Wild Goose | 3,531 | 4,506 | 6,013 | 5,382 | 3,382 | 342 | 18 | 23,174 | 2,687 | 4 | 10,748 |
|    | The X House | 4,086 | 1,660 | 882 | 534 | | | | 7,162 | 498 | 3 | 1,494 |
| P4 | Traditional House | 1,473 | 6,405 | 4,041 | 2,625 | 48 | | | 14,592 | 3,524 | 5 | 17,620 |
|    | Postmodern House | 375 | 412 | 222 | 21 | | | | 1,030 | 534 | 3 | 1,602 |
| P5 | Weatherbeaten Face | 3,072 | 7,264 | 5,161 | 2,141 | 380 | 78 | | 18,096 | 2,012 | 5 | 10,060 |
|    | Lena Face | 3,116 | 2,555 | 2,402 | 2,693 | 625 | 72 | | 11,463 | 376 | 3 | 1,128 |
| P6 | Blue Poles | 44,796 | 24,500 | 8,091 | 3,812 | 288 | | | 81,487 | 9,423 | 6 | 56,538 |
|    | Mona Lisa | 15,248 | 11,249 | 7,849 | 1,637 | 196 | | | 36,179 | 2,673 | 4 | 10,692 |
| P7 | Kodak Mountains | 14,365 | 10,918 | 3,186 | 558 | 174 | 96 | | 29,297 | 2,261 | 5 | 11,305 |
|    | Kodak Island | 2,361 | 4,443 | 3,554 | 1,772 | 629 | 179 | 33 | 12,971 | 1,053 | 3 | 3,159 |
| P8 | Woods | 35,676 | 26,334 | 4,778 | 504 | 207 | | | 67,499 | 7,193 | 6 | 43,158 |
|    | Pond | 16,085 | 7,606 | 1,549 | 732 | 468 | 87 | | 26,527 | 2,816 | 4 | 11,264 |

To further illustrate how the structural beauty was calculated, let us use the two images –weather-beaten face and Lena face – as examples. Table 5 demonstrates the results of the calculation, indicating that the weather-beaten face is more living or more structurally beautiful than Lena face.

Table 5: Comparison of the structural beauty of two images
(Note: This table supplements Table 4 to show an example of how structural beauty (LR) is calculated from substructures at different iterations. I = Iteration, D = Decomposable, S = Substructures, and % = percentage of D to S.)

| | Weather-beaten face | | | | | Lena face | | | |
|---|---|---|---|---|---|---|---|---|---|
| I | D | S | % | LR | I | D | S | % | LR |
| 1 | 1 | 768 | 1.4% | 3,072 | 1 | 1 | 779 | 0.9% | 3,116 |
| 2 | 11 | 1,856 | 1.1% | 7,264 | 2 | 7 | 636 | 3.0% | 2,555 |
| 3 | 20 | 1,206 | 1.8% | 5,161 | 3 | 19 | 646 | 2.3% | 2,402 |
| 4 | 22 | 598 | 1.0% | 2,141 | 4 | 15 | 655 | 1.8% | 2,693 |
| 5 | 6 | 112 | 2.7% | 380 | 5 | 12 | 191 | 1.0% | 625 |
| 6 | 3 | 26 | 0.0% | 78 | 6 | 2 | 24 | 0.0% | 72 |
| Σ | 63 | 4,566 | | 18,096 | Σ | 56 | 2,931 | | 11,463 |

### 4.3 Decomposable substructures and their centroids

If the centroids of the substructures of an image constitute the skeleton of the image, as shown and discussed above, those decomposable substructures and their centroids appear to be the most salient spots of the image. Figure 7 (L1–L8) demonstrates the decomposable substructures (black grounds) and their centroids (red dots). We can note that the decomposable substructures can be seen as sketches that well represent the corresponding images themselves. The red dots look scattered, yet they are the centroids of the decomposable substructures. In fact, the centroids are not scattered at all, and instead they form interconnected wholes or graphs for individual images (G1–G8 of Figure 7). The number of nodes (or substructures) and the number of colors (or hierarchy) of graphs can help differentiate the structural beauty of images. The more substructures, the more beautiful the image, and the higher the hierarchy, the more beautiful the image, as indicated by column V in Table 3.

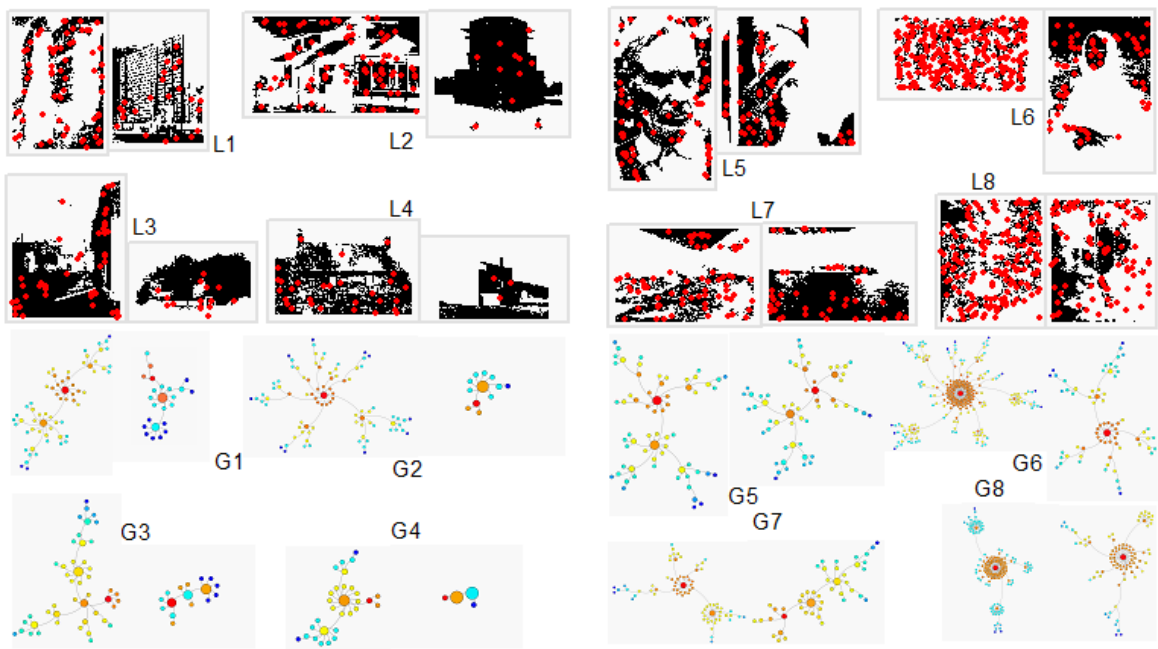

Figure 7: (Color online) Centroids of those decomposable substructures of the images
(Note: The black background represents the decomposable substructures, whose centroids are red dots in L1– L8. The decomposable substructures constitute an interconnected graphs (G1–G8) in which different colors represent different levels of recursion, while dot sizes show the degree of connectivity of both inlinks and outlinks. As the figures show, the contrast between the left and the right for the first four pairs is much greater than that for the second four pairs. This difference is also well reflected in the V value in Table 3.)

It is worth noting that the first four pairs of images are about traditional and modernist buildings, with the traditional ones having a higher degree of livingness. This fact indicates that traditional buildings have far more substructures or far more decomposable substructures for achieving a very steep hierarchy than their modernist counterparts. In other words, modernist buildings deliberately remove substructures to achieve a very flat hierarchy that make them less living or even deadly. This insight is also an important lesson we learned from the classic work (Jacobs 1961) on the notion of vitality in neighborhoods. Another notable pair is that of Blue Poles by Jackson Pollock (1912– 1956) and the Mona Lisa by Leonardo da Vinci (1452–1519). Blue Poles is found to be more living or more structurally beautiful than the Mona Lisa. This is clearly reflected in the pair of graphs (Figure 8), where Panel G6 shows their decomposable substructures. Due to its popularity, Blue Poles was found to be fractal, and was previously studied under the fractal geometry (Taylor et al. 1999). Classic paintings or fractal structures in general have a high degree of structural beauty, yet the structural beauty is for the sake of science (Mandelbrot 1982, 1989, Griffith 2022). Through the notion of structural beauty, we have made it clear not only why a structure is beautiful, but also how beautiful the structure is.

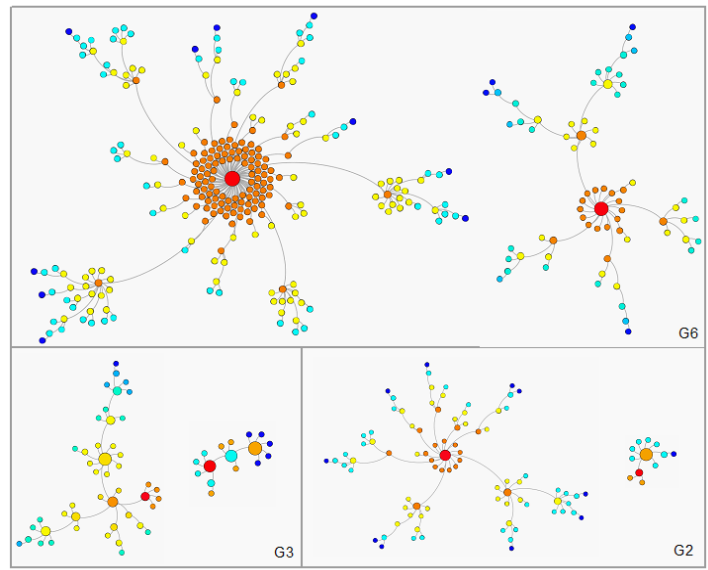

Figure 8: (Color online) Zoom-in view of the three pairs of graphs
(Note: Centroids of the decomposable substructures of an image constitute an interconnected whole or graph. Each graph represents the skeleton of the corresponding image, and its structural beauty can be judged according to the rule: the more nodes, the more beautiful the image, and the higher hierarchy (indicated by colors), the more beautiful the image. There is little doubt for every pair of graphs that the left graph is more living or more beautiful than the right. In addition, the graphs can be perceived as the organized complexity, a key concept in Jacobs (1961), inspired by the work by Weaver (1948).)

The case studies above have shown that images are indeed living, although the degree of livingness varies from one image to another. The recursive approach to structural beauty begins with two subwholes – figure and ground – representing either light or dark side of the image. From either the figure or the ground, the structural beauty can then be calculated respectively for differentiating two images. In other words, the recursive approach can be applied to both the figure and ground sides. This implies that living images are living from both the figure and the ground of images. All of the substructures are derived from the images themselves towards the figure of the figure of the figure and so on. The recursive approach can not only effectively differentiate two images in terms of their livingness, but also help reveal the three major findings outlined above. To supplement the case studies, Appendix A further demonstrates that the same approach applies to georeferenced images as well. The recursive approach is therefore better or more robust than the non-recursive one.

## 5. The livingness of space: Related work, application, and implication
Any space or image has a certain degree of livingness or structural beauty, and it can be sensed in the mind and heart. Conventionally, the degree of livingness was judged by the mirror-of-the-self experiment (Alexander 2002–2005, Wu 2015). That is, two images are put side by side and the human subject is asked to choose the one of the two, to which he/she has a better sense of wholesome feeling. The experiment is intended to capture the subject's genuinely liking from the bottom of his/her heart, so it provides an objective judgement. In other words, the judgement is a kind of human feeling triggered by the underlying living structure rather than idiosyncratic feelings. Recently, eye-tracking and other biometric data (Salingaros, 2020, Lavdas et al. 2021) have been used to capture the kind of objective judgement. Essentially, the livingness of space or living structure in general is a mathematical concept and physical phenomenon, and it can also be reflected in the mind psychologically. Based on the living structure, Salingaros (1997) provided a rough mathematical approximation for computing the degree of livingness of the 24 famous buildings. The line of research can be traced back to the classic work on aesthetic measure (Birkhoff 1933, Eysenck 1942, Douchova 2015) and it continues to be of interest to scientists and scholars; see the recent survey paper (Perc 2020) that has offered a comprehensive overview across a range of the social sciences and humanities.

The livingness of space or structural beauty has significant applications in geography and geographic information science (GIScience). It can help better understand human experience of space such as place attachment and place making (e.g., Tuan 1977, Lewicka 2011, Seamon 2018). This is because the livingness of space or the sense of place is no longer conceived as opinion or personal preference, but a matter of measurable fact. In human geography and GIScience, there is a long history of research interest in humanistic perspectives of geographic space, which requires going beyond the division between science and humanities toward the third culture (Sui 2004, Jiang and Sui 2014, Mennis 2018, Young and Kelly 2017). The recursive approach or the livingness of space in general may offer a new perspective on why natural scenes or living places have healing and nurturing effect on human beings (e.g., Ulrich 1984, Wilson 1984, Kaplan and Kaplan 1989). This of course warrants further research.

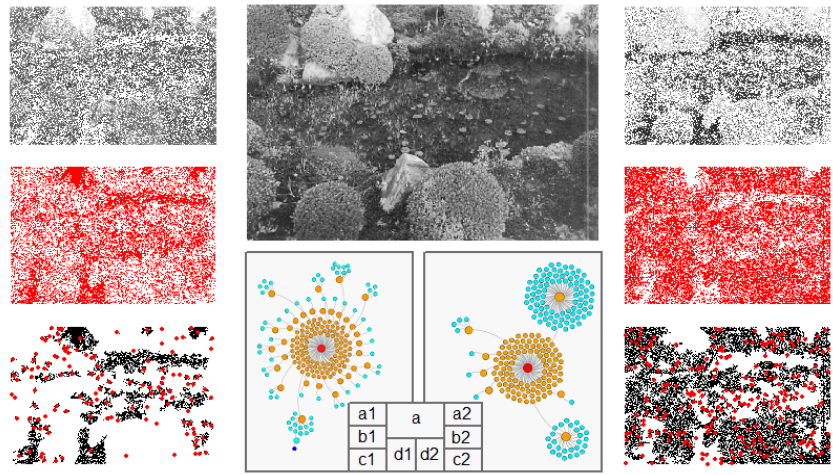

Figure 9: (Color online) The recursive approach to computing the structural beauty of the fishpond (Note: The image fishpond as a whole (a) consists of two subwholes, which are called Figure (a1) and Ground (a2), with which all recursively defined substructures can be generated. The centroids of the substructures are shown in (b1) and (b2), while the decomposable substructures and their centroids are shown in (c1) and (c2). The decomposable substructures constitute interconnected wholes, (d1) and (d2). The corresponding statistics is documented in Table 6.)

This study or the livingness of space in general adds a deep implication on image understanding and computer vision. For example, the morphological skeleton (Umbaugh 2017) and saliency of images (Kadir and Brady 2001, Borji and Itti 2013) can be effectively captured by the centroids of substructures. However, research questions remain regarding how the living structure perspective is comparable to previous approaches to morphological skeleton and saliency maps. While conventional image understanding concentrates on things (objects or features) that human eyes can recognize or natural language can describe with "words" (Deng et al. 2009), the recursive approach to structural beauty captures all substructures that may or may not correspond to human recognizable or language describable things by "words". The auto-generated substructures, and the recursively generated substructures in particular, capture the wholeness of the image. In this connection, the substructures represent a new way of image understanding. While the human recognized objects or features tend to be fragmented as "words", the substructures represent a coherent whole or living structure. To make this point clear, let's examine in detail another living image the fishpond image (Figure 9, Table 6) with which structural beauty is calculated from both the figure and the ground.

Table 6: Statistics about the structural beauty of the image fishpond
(Note: The structural beauty of the image can be calculated, with respect to its two subwholes: the figure and the ground (Figure 9).)

| Image (Fishpond) | | | Figure (Light) | | | Ground (Dark) | | |
|---|---|---|---|---|---|---|---|---|
| Pixel (P) | Number (P) | | 262,062 | | | | | |
| | % (cut value 102) | | 43.4% | | | 56.6% | | |
| Substructure (S) | Number (S) | | 25,718 | | | 26,665 | | |
| | % (S/P) | | 9.8% | | | 10.2% | | |
| Decomposable (D) | Number (D) | | 200 | | | 232 | | |
| | % (D/S) | | 0.8% | | | 0.9% | | |
| | Iteration (I) | | 4 | | | 3 | | |
| Levels of recursion | LR0 | S0 | D0 | 144,452 | 20,636 | 1 | 41,465 | 8,293 | 1 |
| | LR1 | S1 | D1 | 14,809 | 3,477 | 132 | 139,893 | 17,208 | 107 |
| | LR2 | S2 | D2 | 6,569 | 1,592 | 66 | 3,580 | 1,164 | 124 |
| | LR3 | S3 | D3 | 39 | 13 | 1 | | | |
| | Total | Total | Total | 165,869 | 25,718 | 200 | 184,938 | 26,665 | 232 |
| Structural beauty | LR | | 165,869 | | | 184,938 | | |
| | V (D*I) | | 800 | | | 696 | | |

Different people may perceive this image differently, depending on which objects or features they are familiar with or interested in. The fishpond image contains many objects or features, which are termed as centers (Alexander 2002–2005) or substructures. Identifying these substructures is probably the first step of image understanding, but the livingness of image lies on how these individual substructures constitute a coherent whole. The following passage (Alexander 2002–2005) is a typical description of the livingness of the image.

> "Here we have the water, the waves, the fish, the plants, the overhanging bushes, the lilies in the pond, the lily leaves, the mud on the bottom, the caddis, the moss on the rocks, the slime in the water, the flow of water through the pond. Each of these centers, too, is brought to life, and has its life intensified, by the other centers. The fish themselves, for instance, live more intensely when the stream is flowing, so that the dissolved oxygen is constantly replenished. The shade, within the water, formed by the rocks and lily pads, allows the fish a place to cool themselves. The stream flow itself is brought to life, intensified as a center, by the eddies and turbulence at the edge, which make the pond."

A strong sense of livingness is triggered by the fishpond image, and the livingness can be revealed by the underlying substructures. In fact, there are far more substructures (or centers) than what was described in the above passage through words or phrases such as the water, the waves, the fish, and the overhanging bushes. Many of them are beyond what can be described by any natural language. As shown in Figure 9 and Table 6, there are a total of 25,718 and 26,665 substructures from the figure and the ground, respectively. Among the substructures, there are 200 and 232 decomposable substructures. The describable centers in the above passage, such as the water, the waves, the fish, and the overhanging bushes are just a small subset of the most salient of the massive auto-generated substructures. Those describable ones constitute the surface order, while the massive number of substructures constitute a sort of the deep order. It is the deep order that underlies the quality without a name (Alexander 1979). Put simply, the deep order is the quality without a name.

## 6. Conclusion

Any space possesses a certain degree of livingness or structural beauty, although the degree varies from one to another depending on its internal geometry of substructures. Reflected in an image, a substructure is a set of pixels whose pixel values are greater (or less) than the average pixel value. Thus, an image can be viewed as a set of recursively defined substructures, rather than a set of pixels or a set of human recognizable objects as conventionally conceived. The major difference between substructures and objects lies in the fact that substructures are defined by pixels themselves from the bottom up, while objects are defined by human eyes. In this paper, we developed a recursive approach to the structural beauty of images by considering all recursively defined substructures. Through the case studies, we have verified that the recursive approach is better or more robust than the non-recursive approach, although both approaches are based on the same principle: the more substructures than more beautiful, the higher hierarchy of the substructures the more beautiful. We have also verified that the decomposable substructures alone can be used to differentiate two images in terms of their structural beauty, by multiplying the number of decomposable substructures and their levels of recursion. This implies that the more decomposable substructures the more beautiful, and the more levels of recursion the more beautiful.

In addition to the verification of the recursive approach, we have made three major findings. First, the number of substructures of an image is far lower (for example, three percent on average) than the number of pixels, and the centroids of the substructures can capture very well the skeleton or saliency of the image. Second, all the images have the recursive levels more than four, indicating that they are indeed living images. This second finding implies that that any space or matter has a certain degree of livingness or life, according to its internal geometry. Third, no more than 2 percent of the substructures are decomposable, which means that there are far more less-living substructures than more-living ones. To echo the epigraph, we have the following statement about living structure and substructures of an image: in a living structure, every substructure is unique, and the different substructures also cooperate, with no substructures left over, to create a global whole – a whole that can be identified by everyone who is part of it. It is essentially the global whole or wholeness that triggers a sense of livingness in the human mind and heart. The livingness or structural beauty entails that there is a shared notion of livingness among people and even different peoples. It will open a new horizon for research on human experience of space – a sense of places and place attachment – and more importantly on place making. Our future work will concentrate on how the livingness is reflected in the human mind, and whether or how the kind of reflection varies from people to people in terms of their culture, gender, and races.

**Data and code availability statement**
The data used and generated in this study are publicly available at Figshare, which is accessible with the following link https://figshare.com/s/e4a69ee23511de986239. The file package includes (1) an Excel file with results out of the eight pairs of images' analyses, (2) Shapefiles for all pairs both the decomposable substructures and non-decomposable substructures, and (3) all output results of the living structure algorithm for each individual image. The living structure algorithm is based on Python 2.7 (https://www.python.org/download/releases/2.7/) and uses the following libraries: ArcPy (10.8) and Numpy. (The Python scripts process an input image into recursively defined substructures, and into a tree-like network of decomposable substructures.) The software tools used in the study include ArcGIS 10.8 (https://www.esri.com/en-us/arcgis/products/arcgis-desktop/overview) for the processing and analysis of spatial data, Microsoft Excel (https://www.microsoft.com/en-us/microsoft-365/excel) for storing and analyzing numerical results, and Gephi 0.9.2 (https://gephi.org/) for network analysis and visualization.

**Appendix A: Verification of the recursive approach against georeferenced images**

The recursive approach developed in this paper was initially verified by ordinary images that all have the same size and same resolution. Given the same condition, the verification should logically apply to any pair of georeferenced images as well. In this Appendix, we chose two pairs of georeferenced images to demonstrate that how the livingness of space can be well captured (Figure A1). The first pair is about two nighttime images: Belgium, the Netherlands, and Luxburg (Benelux, Panel a1) in contrast to Stockholm region (Panel b1). The central European area including the three countries is much more populated than the Stockholm region, so the former is full of human settlements (far more smalls than larges) therefore more living than the latter; see Panels a2 and b2 or a3 and b3. This fact is clearly reflected in the livingness scores both LR and V.

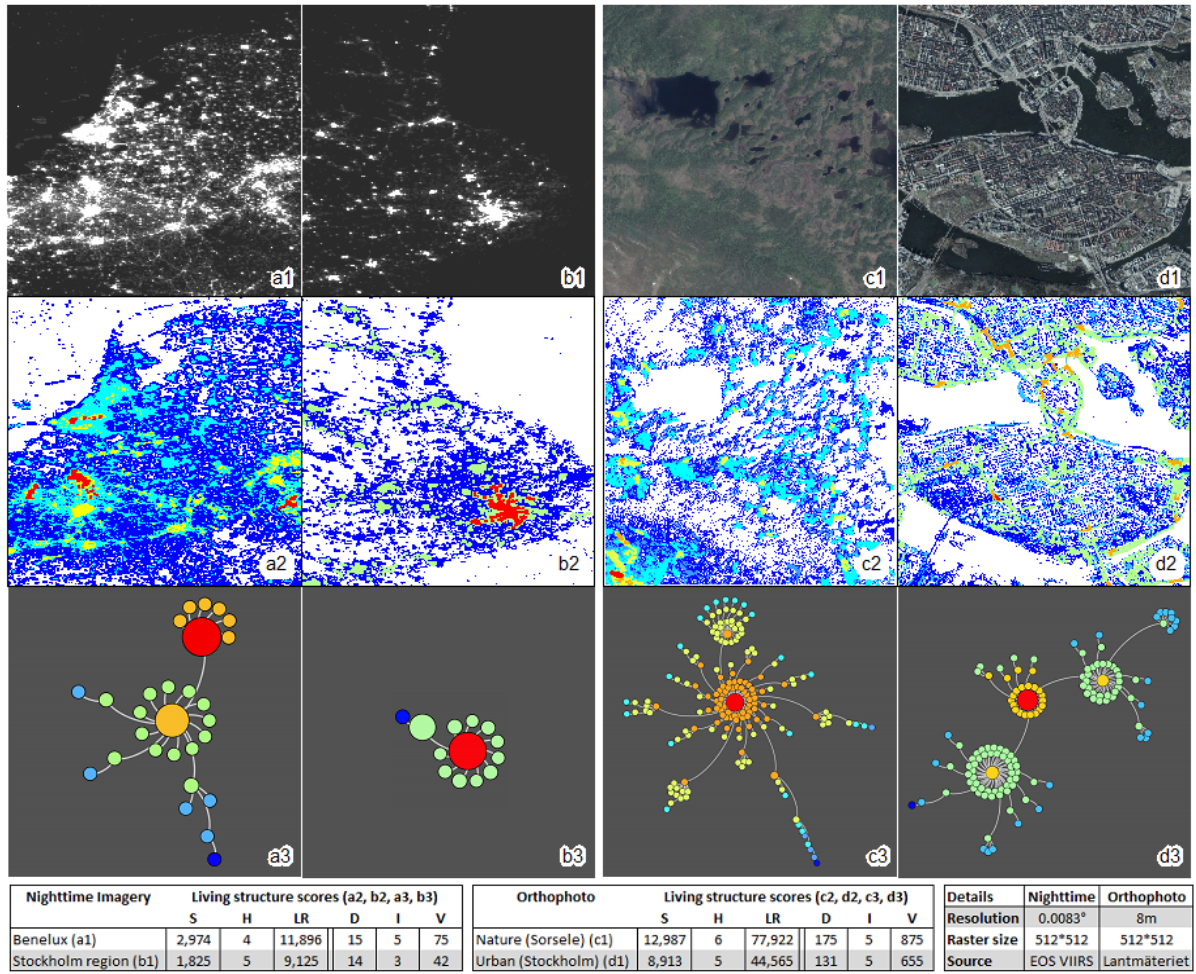

Figure A1 (Color online): The recursive approach applied to the four georeferenced images
(Note: The source images (a1– d1), their corresponding living structures both decomposable and undecomposable (a2– d2), and the decomposable living structures in a graph format (a3– d3). For the living structures shown in the second row (a2– d2), there are numerous first-iteration substructures (in blue), a very few last-iteration substructures (in red), and some in between the first- and last-iteration substructures (in other colors of the spectrum between blue and red). Some of the substructures are nested to each other, with different colors showing the nested relationship, shown in both the second and third rows.)

The second pair is about two satellite images: one from countryside of Sweden (Panel c1) and the other from the center of Stockholm city (Panel d1). Again, the two areas are with the same physical size and their images are with the same resolution. Different from the nighttime images that capture human settlements at night, the satellite images capture geographic features of various kinds. In this case, the natural scene is supposed to be more living than the urban scene (Panels c2 and d2, or c3 and d3). It is indeed true that the natural scene has higher livingness scores than the urban scene.